# New Solar Irradiance Measurements from the *Miniature X-Ray Solar Spectrometer* CubeSat

Thomas N. Woods[1], Amir Caspi[2], Phillip C. Chamberlin[3], Andrew Jones[1], Richard Kohnert[1], James Paul Mason[1], Christopher S. Moore[1], Scott Palo[1], Colden Rouleau[1], Stanley C. Solomon[4], Janet Machol[5], and Rodney Viereck[5]

[1]University of Colorado, Boulder, CO
[2]Southwest Research Institute, Boulder, CO
[3]NASA Goddard Space Flight Center, Greenbelt, MD
[4]National Center for Atmospheric Research, Boulder, CO
[5]NOAA Space Weather Prediction Center, Boulder, CO

**Abstract**

The goal of the *Miniature X-ray Solar Spectrometer* (*MinXSS*) CubeSat is to explore the energy distribution of soft X-ray (SXR) emissions from the quiescent Sun, active regions, and during solar flares, and to model the impact on Earth's ionosphere and thermosphere. The energy emitted in the SXR range (0.1–10 keV) can vary by more than a factor of 100, yet we have limited spectral measurements in the SXRs to accurately quantify the spectral dependence of this variability. The *MinXSS* primary science instrument is an Amptek, Inc. X123 X-ray spectrometer that has an energy range of 0.5–30 keV with a nominal 0.15 keV energy resolution. Two flight models have been built. The first, *MinXSS*-1, has been making science observations since 2016 June 9, and has observed numerous flares, including more than 40 C-class and 7 M-class flares. These SXR spectral measurements have advantages over broadband SXR observations, such as providing the capability to derive multiple-temperature components and elemental abundances of coronal plasma, improved irradiance accuracy, and higher resolution spectral irradiance as input to planetary ionosphere simulations. *MinXSS* spectra obtained during the M5.0 flare on 2016 July 23 highlight these advantages, and indicate how the elemental abundance appears to change from primarily coronal to more photospheric during the flare. MinXSS-1 observations are compared to the *Geostationary Operational Environmental Satellite* (*GOES*) X-Ray Sensor (XRS) measurements of SXR irradiance and estimated corona temperature. Additionally, a suggested improvement to the calibration of the *GOES* XRS data is presented.

**Introduction**

There is a rich history of solar X-ray measurements over the past three decades, but a significant gap remains of accurate soft X-ray (SXR) spectral measurements in the 0.2–3 keV (4–60 Å) range. There were many new discoveries about solar flares during the 1980s with solar SXR spectral measurements from the DoD P78-1 (Doschek 1983), NASA *Solar Maximum Mission* (Acton *et al*. 1980), and JAXA *Hinotori* (Enome 1983) satellites. For example, Doschek (1990) provides results about flare temperatures, electron density, and elemental abundances. Sterling *et al*. (1997) provides a review of flare measurements from the *Yohkoh* Soft X-ray Telescope (Acton *et al*. 1999) and the *Compton Gamma Ray Observatory* (*CGRO*, Fishman *et al*. 1989) for the hard X-ray (HXR) range. These earlier missions laid a solid foundation about flare physics and flare spectral variability that continue with the *Reuven Ramaty High Energy Solar Spectroscopic Imager* (*RHESSI*, Lin *et al*. 2002) and the *Solar Dynamics Observatory* (*SDO*, Pesnell *et al*.

2012) for the HXR (10–1000 keV) and extreme ultraviolet (EUV: 0.01–0.2 keV) ranges, respectively. The most recent solar SXR spectral measurements have been made by *CORONAS-PHOTON* SphinX (Gburek *et al*. 2011; Sylwester *et al*. 2012) and *MErcury Surface, Space ENvironment, Geochemistry, and Ranging* (*MESSENGER*) Solar Assembly for X-rays (SAX) (Schlemm *et al*. 2007; Dennis *et al*. 2015); these have been limited to energies above 1.5 keV (wavelengths below 8 Å) and energy resolutions of 0.4–0.6 keV FWHM. With solar flare spectral variability expected to peak near 0.6 keV (20 Å) (Rodgers *et al*. 2006), we were motivated to develop the *Miniature X-ray Solar Spectrometer* (*MinXSS*) mission (Mason *et al*. 2016) to make new SXR spectral observations for a more complete understanding of flare variability and energetics that fill the wavelength/energy gap between the successful HXR and EUV missions. Examples of the solar SXR spectra are presented here for the M5 flare on 2016 July 23 to highlight the *MinXSS* capability.

Broadband solar SXR measurements have been made nearly continuously for the past four decades, but these measurements cannot directly quantify the varying contributions of emission lines (bound-bound) amongst the thermal radiative recombination (free-bound) and thermal and non-thermal bremsstrahlung (free-free) continua. These solar SXR measurements include two bands over 1.6–25 keV (0.5–8 Å) by the *Geostationary Operational Environmental Satellite* (*GOES)* X-Ray Sensor (XRS; Garcia 1994) since the 1970s and the even broader band of 0.2–12 keV (1–70 Å) from several missions including the *Yohkoh* SXT (1991–2001), *Student Nitric Oxide Experiment* (*SNOE*, 1998–2002; Bailey *et al*. 2000), *Thermosphere-Ionosphere-Mesosphere Energetics and Dynamics* (*TIMED*, 2002–present; Woods *et al*. 2005a), the *Solar Radiation and Climate Experiment* (*SORCE*, 2003–present; Woods *et al*. 2005b), and the *SDO* EUV Variability Experiment (EVE, 2010–present; Woods *et al*. 2012; Didkovsky *et al*. 2012). These broadband SXR measurements have been helpful for resolving differences in terrestrial ionosphere models and measurements, but there remain differences in the understanding of the solar SXR spectral distribution and atmospheric photoelectron flux. There are also large differences, by up to a factor of three, between the irradiance levels measured by these broadband SXR photometers. The lack of spectral resolution in the SXR range is thought to be the culprit for most of these disagreements, primarily because the conversion from the instrumental measurement to solar irradiance requires assumptions about the poorly known shape and normalization of the incident solar SXR spectrum. The new *MinXSS* spectral irradiance measurements can now be used to address those differences; one example is shown here for the *GOES* XRS measurement.

The *MinXSS* primary instrument is an X-ray spectrometer – a commercial unit from Amptek, Inc. called the X123 Silicon Drift Detector (SDD). The X123 SDD has an active area of 25 mm$^2$, an effective Si thickness of 0.5 mm, a nominal 8 µm-thick Be filter on the detector vacuum housing, an active 2-stage thermoelectric cooler (TEC) on the detector, and sophisticated multichannel analyzer (MCA) detector electronics. The thickness of the Si determines the high energy (short wavelength) sensitivity limit, and the thickness of the Be filter sets the low energy (long wavelength) limit. This X-ray spectrometer measures individual photons with energies from ~0.5 keV to 30 keV (0.4 Å to 25 Å). Its energy resolution is nominally ~0.15 keV FWHM, with an energy bin size of ~0.03 keV (Moore *et al*. 2016). Spectra are accumulated with a nominal 10 s cadence. With the addition of a pinhole aperture to limit the solar flux during large flares and the fairly large correction for photoelectrons from the Be filter for the lowest energy photons, the solar full-disk measurements from *MinXSS*-1 X123 are most accurate between ~0.8 keV and 10 keV (1.2 Å to 16 Å). With more analysis, the energy range could be extended for the *MinXSS*-1 X123 spectra, but in this paper, only measurements in the 0.8–10 keV range are discussed. In addition, *MinXSS*-2 – expected to launch in 2017 April – has an upgraded version of the X123 called the Fast SDD, which has lower noise and wider dynamic range for enhanced measurements. Mason *et al.* (2016) provide more details about the *MinXSS* project and spacecraft design, including an overview of the Blue Canyon Technologies (BCT) attitude determination and control system (ADCS) that provides about 10″ stability for the *MinXSS* solar observations.

*MinXSS* was deployed from the International Space Station on 2016 May 16, and began normal operations on 2016 June 9. The Sun was relatively quiet in June, more active in July, and quiet again in August and September. So far, *MinXSS*-1 has observed more than 40 C-class flares and 7 M-class flares. An example of the *MinXSS* observations for one of these M-class flares is presented, followed by comparison to *GOES* XRS data over the *MinXSS*-1 mission.

### M5 Flare Example

The largest flares so far observed by *MinXSS* occurred on 2016 July 23 with an M5.0 flare that peaked at 2:11 UT, an M7.6 flare that peaked at about 5:16 UT, and an M5.5 flare that peaked a few minutes later at 5:31 UT. The X123 pinhole aperture was sized for linear response up to about M5 and then detector dead time and spectral pulse pile-up corrections are required for larger flares. As these corrections are not yet applied in *MinXSS* data processing, data with values above the *GOES* M2 level have been excluded in this analysis. We note that the *MinXSS*-2 X123 Fast SDD has four times wider dynamic range for the same size pinhole aperture, and its linear response is expected to be linear up to X-class flares.

The *MinXSS*-1 pre-flare and flare SXR spectra for the M5.0 flare are shown in Figure 1A, and the total counts for each X123 spectrum are plotted as a time series in Figure 1B. The X123 Level 1 data product includes 1-min averages for improved statistics compared to the nominal 10-sec observational cadence. For this figure, the pre-flare spectrum is a 21-minute average centered at 01:36 UT, and the flare spectrum is a 9-minute average centered at 02:00 UT. The grey regions in Figure 1B are when *MinXSS* was in orbit eclipse. The *GOES* XRS-B (1–8 Å) data are also included in Figure 1B.

The SXR irradiance for the flare spectrum increased, relative to the pre-flare level, by a factor of ~10 near 1 keV, and by more than a factor of 100 for SXR energies greater than 3 keV. This large increase at the higher energies is an indicator of a hotter plasma temperature during the flare. In addition, the hot coronal Ca and Fe lines above 3 keV become more obvious during the flare, as the additional hot plasma leads to brighter emission in these lines. These emission lines are from elements with low first ionization potential (FIP) and are sensitive to the "FIP effect" where the elemental abundance, relative to photospheric composition, of low-FIP elements is often observed to be enhanced by a factor of about 2–4 for coronal closed magnetic field features and closer to 1 (photospheric) for open field features (Laming 2015). Consequently, the abundance enhancement factor, as measured by emission lines, is a useful diagnostic for exploring the origins of the hot coronal plasma, and thus coronal heating processes and composition.

For *MinXSS* data studies, the X123 spectra provide important abundance information that broadband SXR measurements, such as from *GOES* XRS, cannot provide. Phillips (2004) discusses the importance of Ca XIX, Fe XXV, and Ni XXVII lines in the 3.8-10 keV range for studying abundance in flare spectra and how a lower abundance likely indicates an enhanced chromospheric contribution to the hot coronal plasma. X123 fully covers this 3.8-10 keV range. There have been interesting, but sometimes conflicting, results for flare abundance factors. For example, Sylwester *et al.* (1984) report different Ca abundance factors for different flares with *SMM* observations, Schmelz (1993) reports high corona-like abundance factors for flares measured with *SMM*, and Warren (2014) reports low photosphere-like abundance factors during flares with *SDO* EUV data. Additionally, Fludra & Schmelz (1999) and Dennis *et al*. (2015) indicate a mixture of coronal and photospheric abundance factors for different elements during flares observed by *Yohkoh* and *MESSENGER* SAX, respectively.

The abundance factors for the X123 observations are derived through fitting CHIANTI spectra (version 8.0.5 code, version 8.0.2 database, Dere *et al.* 1997; Landi *et al.* 2013; Del Zanna *et al*. 2015) to the X123 spectra with two temperature components. The first is a "hot" component using the higher energy range of the X123 spectra. If the X123 spectrum corrected by the first temperature model spectrum has residual flux at the lower energies, then a second cooler component is also fit. The flare spectra all required two temperature components with one temperature being more than 10 MK and another at a few MK, and some of the quiescent spectra only needed a one-temperature component. As shown in the next section, there is a floor of about 2 MK for observed coronal temperatures based on these X123 spectral fits.

These model fits are illustrated in Figure 2 for the pre-flare and flare spectra associated with the M5.0 flare. Caspi *et al*. (2015b) describe similar two-temperature modeling and results with rocket-based X123 measurements in 2012 and 2013, albeit for quiescent emission. These two-temperature component model spectra fit the X123 observations well. These will be compared in the future with more detailed differential emission measure (DEM) model fits by combining X123 SXR, *SDO* EUV and *RHESSI* HXR measurements (e.g., using the DEM techniques by Caspi *et al*., 2014b) Each temperature component is not isothermal but is instead a convolution of a few CHIANTI isothermal spectra to achieve an emission measure (EM) with a $\log_{10}$ temperature (logT) FWHM width of 0.2. The EM distribution of these fits are shown in Figure 2C. Based on the fit and pre-flight calibration uncertainties, the uncertainty of the central logT value is about 0.05, and the EM relative uncertainty is estimated to be about 15%.

The Si, Ca, and Fe emission lines in the SXR range are very sensitive to the abundance value for logT between 6.8 (~6 MK) and 7.7 (~50 MK). The abundance factor is derived for the hot component when logT≥6.8 using the Fe XXV (~6.7 keV; cf. Phillips 2004) emission, and its fit uncertainty is estimated to be about 30%. The abundance factor for the cooler component is assumed to be the same value as for the hotter component. For fitting the first (hot) temperature components that are cooler than logT=6.8, the coronal abundance factor is assumed for those model fits. The coronal abundances given by Schmelz *et al*. (2012) and photospheric abundances given by Caffau *et al*. (2011) were used in the CHIANTI spectra to make the temperature component reference spectra for these model fits. The coronal / photospheric ratio of the low-FIP abundance is 2.183 in these reference spectra.

The derived abundance factors from the X123 spectra are shown in Figure 3B, which indicate a transition to photosphere-like abundance during the flare. This result is consistent with the standard flare model that predicts the majority of hot, SXR-emitting plasma to originate from the chromosphere as a result of "evaporation" from intense heating by non-thermal, accelerated electron beams and by thermal conduction fronts from the corona (Holman *et al*. 2011). In particular, the transition to photospheric abundance for the flare spectrum suggests that a significant source of the hot flaring plasma is from the lower atmosphere (chromosphere), transported to the corona via "evaporation" and mixed with the original plasma in the coronal loops where magnetic reconnection is assumed to have initiated the flare event. Prior studies (e.g., Caspi & Lin 2010; Liu e*t al*. 2013; Caspi *et al*. 2015a) have suggested that significant heating also occurs directly in the corona, which would be consistent with a "hybrid" abundance factor that is between the photospheric and coronal factors. X123's 10-s cadence will be useful to examine the timing of the abundance transition from coronal to photospheric in all of the various emissions (including Mg, Si, S, and Ar, in addition to Fe as discussed here), and will be the subject of future studies.

In addition to the abundance information from the X123 spectra, another important advantage of spectrally resolved measurements is the capability to perform multiple-temperature analysis. In the model fits here, two temperature components are fit, and those temperatures are shown in Figure 3A as the diamond symbols. The single temperature estimate from the ratio of the *GOES* XRS A and B bands (Woods *et al.*, 2008) is also shown in Figure 3A. The X123 hot (Temp-1) temperature component agrees well with the *GOES* XRS estimated temperature. The X123 cooler (Temp-2) temperature component is most important for improved model fits for flare spectra, and this cooler temperature is similar to the pre-flare temperature. For non-flare data, the second temperature component is sometimes not needed for a good model fit or is a slightly cooler than the first fitted temperature.

### *GOES* Calibration Example

Because the *MinXSS* instruments had pre-flight calibrations at the National Institute of Standards and Technology (NIST) Synchrotron Ultraviolet Radiation Facility III (SURF-III; Arp *et al.* 2000) with an accuracy of about 10% (Moore *et al.* 2016), the *MinXSS*-1 solar observations can provide a spectral calibration for the broadband SXR photometers currently operating on *GOES*, *TIMED*, *SORCE*, and *SDO*. A comparison of *MinXSS*-1 results to *GOES* XRS-B is presented here. The NOAA Space Weather Prediction Center (SWPC) analysis of the *GOES* XRS data indicates that the XRS-B band reported values need to be multiplied by 1.43 (divided by 0.70) to convert from flare levels (e.g., C, M, X) to physical irradiance units of W m$^{-2}$. (see http://ngdc.noaa.gov/stp/satellite/goes/doc/GOES_XRS_readme.pdf ). However, the true corrections for the *GOES* XRS reported irradiances are not expected to be a simple scalar factor applicable to all flare phases and magnitudes because the *GOES* XRS irradiance data processing is based on an assumed fixed quiet-sun solar SXR spectrum at 2 MK (Garcia, 1994; Neupert, 2011), while the true solar spectrum is known to vary during and between flares (e.g., Feldman *et al.* 1996; Caspi & Lin 2010; Fletcher *et al.* 2011; Caspi *et al.* 2014a; Warmuth & Mann 2016). In other words, the SXR spectral distribution changes dramatically as the coronal temperature changes (as shown with *MinXSS* data in Figure 1), so the scaling of the irradiance for a broadband SXR photometer will vary with the assumed SXR spectrum (itself dependent on an assumed coronal plasma temperature, density and abundance).

With the new *MinXSS*-1 observations, the relationship between the reported irradiance from *GOES* XRS, and the irradiance obtained by integrating the X123 spectra from 1 to 8 Å, can provide an improved calibration for the *GOES* XRS-B irradiances. This comparison and new suggested calibration result are shown in Figure 4. Panel 4A shows the time series comparison that indicates higher values for the X123 integrated irradiance than for the reported XRS-B values during flares and much lower irradiance for quiescent activity. In Figure 4A, the GOES XRS and X123 data are both plotted with 1-min cadence. The GOES XRS data have no gaps. The X123 data have several gaps during the ~30-min eclipse period every orbit and on a few days when recovering from spacecraft anomalies (e.g., bad ephemeris data on DOY 2016/220). Figure 4B is the ratio of the X123 integrated irradiance to the reported XRS-B irradiance versus the XRS-B level. As expected, this ratio is not a constant value at the low levels because the XRS background level correction is a constant instead of a time-varying correction and at the higher levels because the coronal temperature is changing with the SXR variability as illustrated in Figure 4C. The flattening of this irradiance ratio above about C5 is likely related to the hot component becoming completely dominant in the 1–8 Å range as seen by both instruments, compared to the cooler few-MK "ambient" component that X123 clearly sees but which is at the limit of the *GOES* XRS-B temperature response. The X123 dominant hot temperature component is plotted in this Panel 4C along with the estimated temperature derived using the ratio of the *GOES* XRS-A (0.5–4 Å) to XRS-B intensities. The *GOES*-derived temperature deviates from the X123-derived temperature below the B7 level; this is related to the XRS-A signal approaching its signal floor and then forcing a bias towards a hotter temperature. In other words, the XRS-A signal is not accurate at the lower levels, thus the temperature derived with the ratio of XRS-A to XRS-B is not accurate below the B7 level (cf. Ryan *et al*. 2012). The X123-derived temperature trend, the green line in Figure 4C, does not have this limitation, and it indicates a temperature transition from a non-flare temperature of 1.8 MK below the B2 intensity to the hot corona flare temperature that reaches ~12 MK around the M1 intensity. There are only a few X123 results above M1, so the appearance of flattening at ~12 MK may not be realistic. Instead the temperature trend likely continues upward for higher *GOES* classes (cf. Feldman *et al*. 1996; Caspi *et al*. 2014a).

The suggested calibration function for scaling the reported *GOES* XRS-B flare levels to obtain physical irradiance units is provided in Figure 4B. This function is almost a constant value of about 1.5 for above the C1 level, consistent with the prior SWPC analysis of the *GOES* calibration. This function deviates to a smaller factor below the C1 level and is about a factor of one at the B2 level where the temperature is estimated to be 2 MK (Figure 4C), which notably is the temperature used for the *GOES* XRS reference spectrum. This calibration function is valid over the measured intensity range of A5 to M2. From initial analysis by NOAA SWPC, this ratio of X123 to XRS-B will be flatter when a revised time-dependent background correction is applied to the XRS-B data. Future *MinXSS* measurements could extend this calibration over a broader flare intensity range. For now, we suggest a constant factor of 1.5 could be used for flare magnitudes above M2. The uncertainty for this calibration function (fitted ratio) is estimated to be 0.15 (or 12% as relative uncertainty) based on the function fit (5%) and X123 pre-flight calibration (10%) uncertainties.

**Conclusions and Future Work**

The *MinXSS*-1 data are providing new information for exploring the highly variable solar SXR spectra. This single example of flare observations by *MinXSS*-1 X123 illustrates the advantages of the SXR spectra over broadband SXR measurements. In particular, multiple-temperature component models can be derived with the SXR spectra; these model results indicate (1) a lower limit of the X123-observed corona temperature at 1.8 MK, (2) typical pre-flare corona temperature of a few MK, (3) typical flare corona temperatures hotter than 10 MK, and (4) elemental composition transitioning from coronal abundance before a flare to photospheric abundance during a flare.

Comparisons to other SXR photometer data sets and analyses of more flare events are in progress and are needed to draw definitive conclusions regarding flare heating processes, temporal evolution, and elemental abundances. More detailed results for flare energetics and active region evolution are expected when combining *MinXSS* SXR spectra with simultaneous *RHESSI* HXR spectra and *SDO* EUV spectra (cf. Caspi *et al*. 2014b). These new SXR spectra are also important for updating solar irradiance spectral models and for using such results as input for terrestrial and planetary atmosphere/ionosphere models to better understand the impacts of the SXR variability. In addition, we are initiating studies of active region evolution using the *MinXSS* data in combination with *Hinode* X-Ray Telescope (XRT) SXR images.

The new *MinXSS-1* calibrated SXR spectra have been used to provide a new calibration for the *GOES* XRS-B data. We are also working on new calibrations for the *GOES* XRS-A band (0.5–4 Å) as well as for the SXR photometers aboard *TIMED*, *SORCE*, and *SDO*. The NOAA National Centers for Environmental Information plan to put out a new GOES XRS data set with physical units using these new *MinXSS*-derived corrections and with a time-varying XRS background. Woods *et al.* (2008) present the technique of using CHIANTI reference spectra in processing the *TIMED* and *SORCE* SXR photometers, and we are working towards replacing these CHIANTI spectra with SXR reference spectra from *MinXSS*. We anticipate these new spectra will resolve the large differences, sometimes as large as factor of three, that currently exist between irradiance measurements by the *TIMED*, *SORCE*, and *SDO* photometers.

**Acknowledgments**

We thank the many students and staff at the University of Colorado for the development, testing, and operations of the *MinXSS* CubeSat. The early student project design classes were supported by NSF. NASA grant NNX14AN84G has supported the flight build, testing, and operations for the *MinXSS* mission.

**Data Source**

The *MinXSS* Level 1 solar SXR spectral irradiance with up to 1-minute cadence, along with user guide and data plotting examples in IDL and Python, are available from the *MinXSS* web site at http://lasp.colorado.edu/home/minxss/. The *GOES* XRS data are available from http://www.ngdc.noaa.gov/stp/satellite/goes/dataaccess.html.

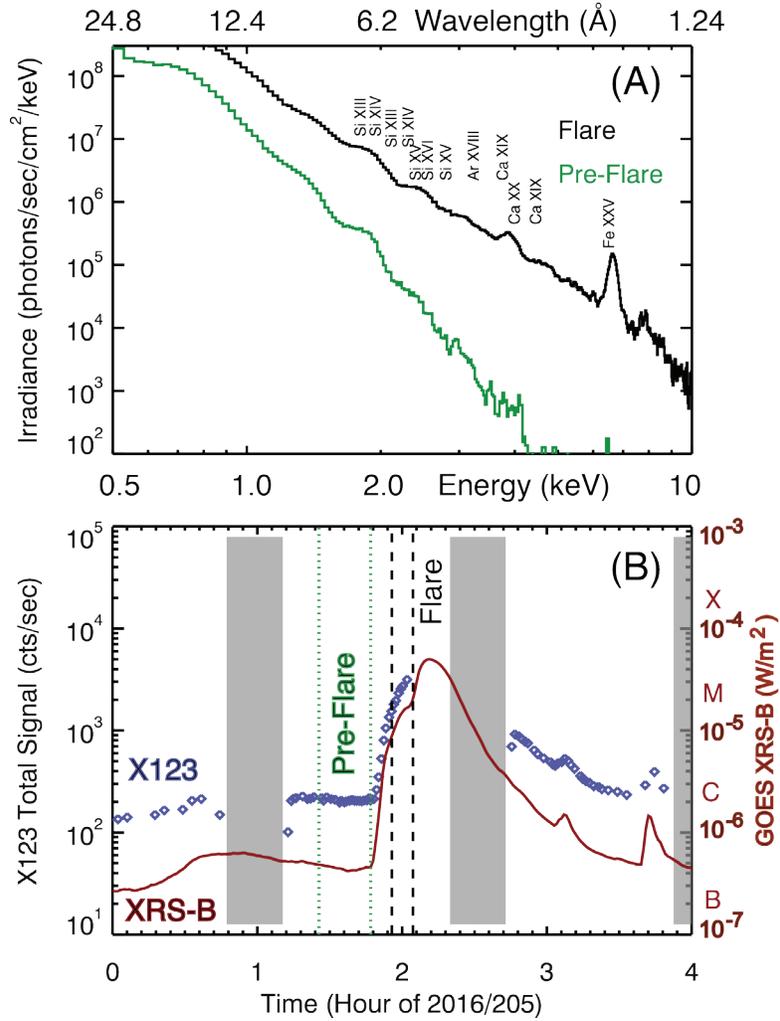

**Figure 1.** Example *MinXSS* measurement of the M5.0 flare on 2016 July 23. (A) Pre-flare (green) and flare (black) spectral irradiance measurements. (B) The *MinXSS* observations (blue diamonds) track well with the scaled *GOES* XRS data (red). The dotted and dashed vertical lines indicate the time range for the pre-flare and flare spectra, respectively. The pre-flare spectrum is a 21-minute average centered at 01:36 UT, and the flare spectrum is a 9-minute average centered at 02:00 UT.

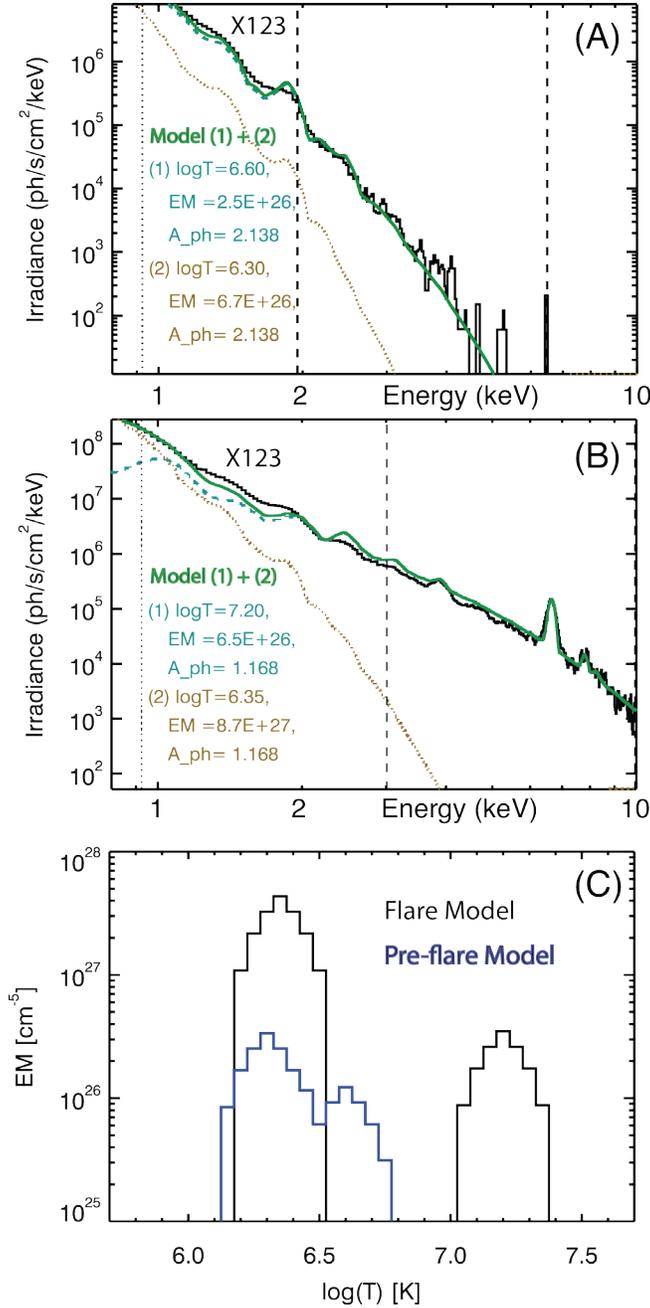

**Figure 2.** Two temperature model fits for the X123 spectra during the M5.0 flare. (A) Pre-flare spectrum has a warm coronal plasma of 2–4 MK with coronal abundances (A_ph is the abundance enhancement factor relative to photospheric). (B) Flare rising-phase spectrum has a brighter warm coronal contribution, like the pre-flare spectrum, plus a hot contribution near 15 MK and with photospheric abundances. (C) The emission measure (EM) for the two temperature contributions are shown for the pre-flare (blue) and flare (black) model fits. In panels A and B, the *MinXSS* spectra are the black lines, the model fits are the green lines, the temperature-1 contribution is light blue, and the temperature-2 contribution is gold. The dashed and dotted vertical lines denote where the temperature 1 and 2 contributions are fitted, respectively.

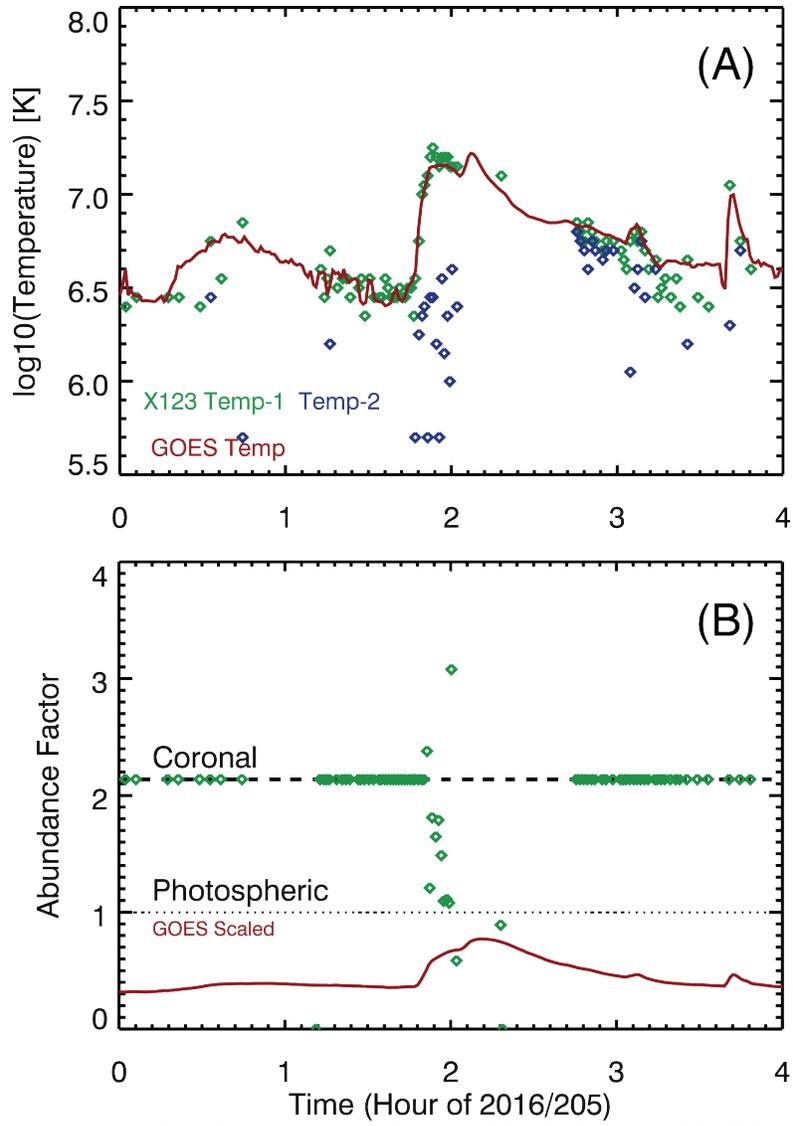

**Figure 3**. Temperature and abundance results for the M5.0 flare on 2016 July 23. (A) Coronal plasma temperatures are estimated from the ratio of the *GOES* XRS two bands (red) and with fitting two temperature contributions to the X123 spectra (green and blue diamonds). (B) The abundance factors estimated from the X123 spectra indicate a significant chromospheric contribution to the hot plasma during the flare.

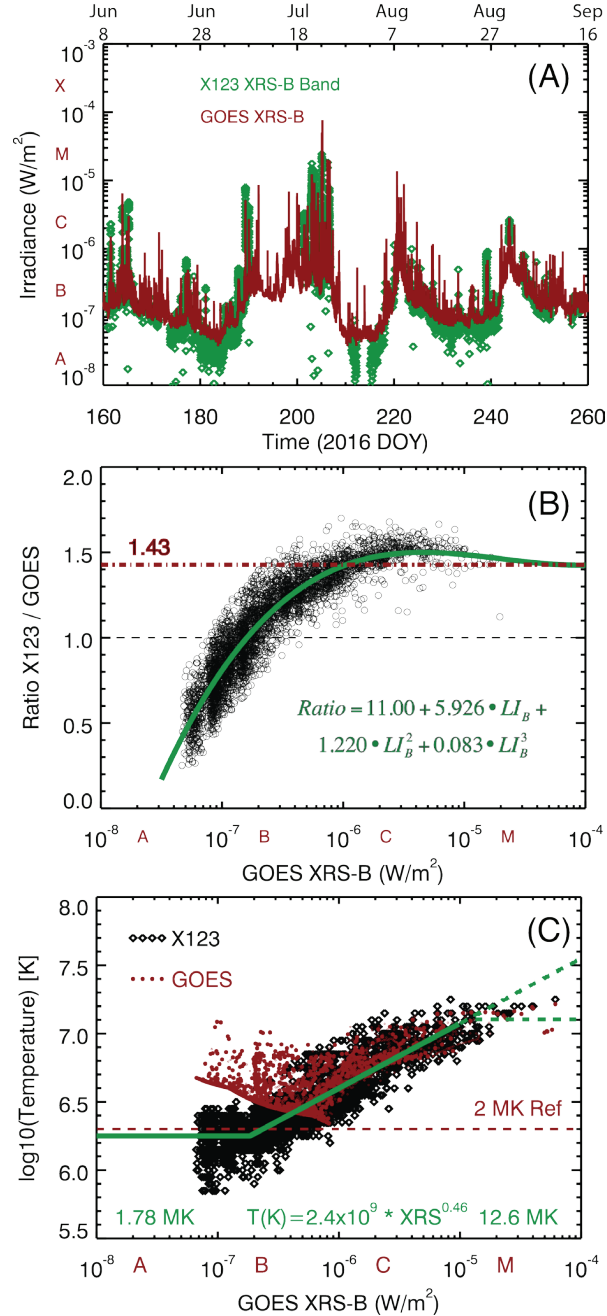

**Figure 4**. Comparison of *GOES* XRS with *MinXSS* data. (A) The *MinXSS* X123 spectra integrated over the 1–8 Å band (green) is compared to the *GOES* XRS-B band on its reported scale (red). (B) The ratio of this X123 integrated band to *GOES* XRS-B is plotted as function of XRS intensity. The red dot-dash line at 1.43 is the NOAA previous irradiance conversion factor for *GOES* XRS-B. Suggestion for new calibration for *GOES* XRS-B is shown as the green line with the function parameters listed in panel B with LI being the log10(*GOES* XRS-B). This calibration function is valid over the intensity range of A5 to M2. (C) The temperature estimated as a function of *GOES* XRS-B intensity. The black diamonds are the hot temperature component for the *MinXSS* SXR spectra, and the red dots are the temperature derived from the *GOES* XRS two bands. Temperature trend fits are shown as the solid green lines, and possibilities for the temperature trend for above the M1 level are shown as the dashed green lines.